\documentclass[10pt]{article}
\usepackage{bbm}
\usepackage[final]{graphics}
\usepackage{amsmath}
\usepackage{amsfonts,amsbsy}
\usepackage{amssymb}
\usepackage{graphicx}

\catcode`\@=11

\newcount\@tempcntc
\def\@citex[#1]#2{\if@filesw\immediate\write\@auxout{\string\citation{#2}}\fi
  \@tempcnta\z@\@tempcntb\m@ne\def\@citea{}\@cite{%
        \@for\@citeb:=#2\do%
    {\@ifundefined{b@\@citeb}%
        {\@citeo\@tempcntb\m@ne\@citea%
                \def\@citea{,\penalty\@m\ }{\bf ?}\@warning%
                {Citation `\@citeb' on page \thepage \space undefined}}%
        {\setbox\z@\hbox{\global\@tempcntc0\csname b@\@citeb\endcsname\relax}
     \ifnum\@tempcntc=\z@ \@citeo\@tempcntb\m@ne%
       \@citea\def\@citea{,\penalty\@m}%
       \hbox{\csname b@\@citeb\endcsname}%
     \else%
      \advance\@tempcntb\@ne%
      \ifnum\@tempcntb=\@tempcntc%
      \else\advance\@tempcntb\m@ne\@citeo%
      \@tempcnta\@tempcntc\@tempcntb\@tempcntc\fi\fi}}\@citeo}{#1}}%
 
\def\@citeo{\ifnum\@tempcnta>\@tempcntb\else\@citea
  \def\@citea{,\penalty\@m}%
  \ifnum\@tempcnta=\@tempcntb\the\@tempcnta\else
   {\advance\@tempcnta\@ne\ifnum\@tempcnta=\@tempcntb \else
\def\@citea{--}\fi
    \advance\@tempcnta\m@ne\the\@tempcnta\@citea\the\@tempcntb}\fi\fi}

\catcode`\@=12

\begin{document}

\title{\bf  Quarkyonic Matter and the Phase Diagram of QCD}
\author{Larry McLerran}
\maketitle
\begin{center}

 Physics Department and Riken Brookhaven Research Center, Building 510A\\ Brookhaven National Laboratory,
  Upton, NY-11973, USA
\end{center}

\begin{abstract}
Quarkyonic matter is a new phase of QCD at finite temperature and density which is distinct from the confined and de-confined phases.
Its existence is unambiguously argued in the large numbers of colors limit, $N_c \rightarrow \infty$, of QCD.  Hints of its existence for QCD, $N_c = 3$, are shown in lattice Monte-Carlo data
and in heavy ion experiments. 
\end{abstract}

\section{Quarkyonic Matter and the Large $N_c$ limit of QCD}

The large $N_c$ limit of QCD has provided numerous insights into the structure of strongly interacting
matter, both in vacuum,\cite{thooft}, and at finite temperature.\cite{thorn}  The large $N_c$ 
approximation allows a correct reproduction of both qualitative and semi-quantitative features of QCD.
If the number of fermions is held finite as $N_c \rightarrow \infty$, then QCD in this limit is confining
in vacuum.   The spectrum of the confined world consists of non-interacting confined mesons and glueballs.   At finite temperature, there is a first order phase transition between a confined world
of glueballs and mesons and an unconfined world of gluons.  The energy density, pressure and entropy
are parametrically of order one in the confined world since the confined states are colorless, but in the de-confined world are of order $N_c^2$,
corresponding to the $N_c^2$ gluon degrees of freedom.  The latent heat of the phase transition is of order $N_c^2$. 

As one approaches the phase transition from lower temperature, the transition is hinted at by the existence of a Hagedorn spectrum of particles, whose density accumulates as one approaches the phase transition temperature.  This accumulation resolves the paradoxical situation that at large $N_c$, hadrons do not interact, which seems to contradict the existence of a de-confining phase transition.  If there is an accumulation of states at the Hagedorn temperature, then the the high density of states very near to the transition temperature can compensate for weak interaction strength, resulting in a strongly interacting gas of hadrons.

The results of lattice computations for QCD  ( $N_c = 3$) are in qualitative accord with the results at large $N_c$.\cite{karsch}  There is a rapid change at at a well defined temperature, although not a strict discontinuity
when a finite number of quark flavors are included.  The presence of quarks makes the order parameter for confinement not so well defined, since the order parameter is $e^{-\beta F_q}$ where inverse temperature is $\beta$ and the change in the free energy of the system, $F_q$, is that due to the addition of a quark.  In a theory with no light quarks, one can still probe the system with a heavy quark source, which corresponds to the order parameter.   In such a theory, $e^{-\beta F_q} = 0$ in the confined phase because the quarks have infinite mass.  In the de-confined phase, the order parameter is finite. When light quarks are included, these light quarks can form bound states with the heavy quark probes, and so the free energy need never be infinite.  The presence of light quarks therefore does not allow for an order parameter for confinement,  In QCD, since there is no order parameter associated with confinement,
there need be no strict phase transition, and there appears not to be for realistic quark masses. 
There is nevertheless a quite rapid transition at a temperature of about $200~MeV$, where the energy density changes by of order $N_c^2$, in accord with large $N_c$ arguments.

The popular wisdom for QCD at finite temperature and density is that there is a line of  cross overs, perhaps converting to a first order phase transition at high density and low temperature, that separates the confined and de-confined world.  Typically plots are made as a function of temperature $T$ and baryon chemical potential, $\mu$. Such a hypothetical diagram is shown in Figure 1.
At very high temperature and density, there may be phase transitions associated with color superconductivity, which affect transport properties of quark-gluon matter, but are not so important for bulk properties such as pressure and energy density.\cite{alford} 
\begin{figure}[htbp]
\begin{center}
\includegraphics[width=0.60\textwidth]{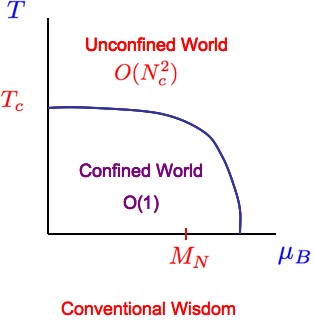} 
\end{center}
\caption{ The conventional wisdom about the QCD phase transition at finite temperature and density.
}
\end{figure}

Unfortunately, the conventional wisdom about the phase diagram has never been explicitly verified.
It is quite difficult to disentangle the assumptions built into various model computations from the
hard predictions of QCD.\cite{fukushima}  Lattice Monte-Carlo computations are extremely difficult at high baryon density.\cite{karsch}

In this talk, I will review the recent results concerning the phase diagram at finite $T$ and $\mu$ in the limit of a large number of colors.\cite{pisarski} In this limit, it is possible to extract model independent results.  The surprise result of these considerations is that in addition to the confined and de-confined phases
of  QCD, there is a third phase.  It turns out that the pressure and energy density of this phase behave like a gas of quarks at very high baryon density, but nevertheless is confined.  Confinement is important for properties of the matter near the Fermi surface, where excitations are required to be bound into color singlets
Rob Pisarski and I named this phase the quarkyonic phase, since it has properties of both high density
baryonic matter and de-confined quark matter.

To understand how such a new phase of matter might come about, we need to understand that dynamical quarks
do not modify the potential between heavy test quarks at large $N_c$.  We shall first consider the case that $N_c \rightarrow \infty$ but that the number of quark flavors is held fixed.  We will later turn to the case where $N_F/N_c$ is held fixed.  This limit with finite number of flavors is easiest to consider since
there is a confined and a de-confined phase.  For finite $N_F/N_c$, there is no distinction between a confined and de-confined phase, although there might perhaps be a remnant of these phases associated with a cross over.  In Figure 2, the gluon loop and quark loop modifications of the potential are shown.  At finite temperature, the first diagram Debye screens the potential at large distances.  This can short out the linear potential when the temperaute is high enough.  The second diagram corresponds to a quark loop and is suppressed by $1/N_c$ at large $N_c$.  When expressed in terms of the t'Hooft coupling, the temperature $T$ and the quark chemical pottential $\mu_Q$, it is of order $\alpha T^2 F(\mu_Q/T) /N_c$  Note that the baryon chemical potential is $\mu_B = N_c \mu_Q$.  Typically, the baryon chemical potential is of order 
the baryon mass $M_B \sim N_c \Lambda_{QCD}$.  The high density limit is $\mu_Q >> \Lambda_{QCD}$.  In order for the quark loops to be important, the quark chemical potential must be of order $\mu_Q \sim \sqrt{N_c} \Lambda_{QCD}$ which approaches infinity in the large $N_c$ limit
\begin{figure}[htbp]
\begin{center}
\includegraphics[width=0.99\textwidth]{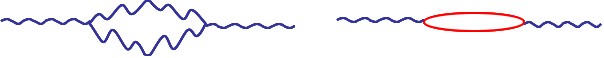} 
\end{center}
\caption{ Diagrams which modify the potential at high energy density .
}
\end{figure}

This argument shows that the presence of baryons at finite density does not affect the values of the confinement-deconfinement transition temperature.  How can there be any non-trivial physics due
to finite baryon number density?  It turns out that at large $N_c$ there is another order parameter,
which is the baryon number density.  Remember that the baryon number density is
\begin{equation}
       \rho_B \sim e^{(\mu_B -M_B)/T} 
\end{equation}
Since both $\mu_B$ and $M_B$ are of order $N_c$, $\rho_B \sim e^{-\kappa N_c}$ so long as
$\mu_B \le M_B$.  Here $\kappa$ is a constant of order 1.  Therefore there is a finite region in the $\mu_B - T$ plane where the baryon number
density is zero.  This is the confined-baryonless world.  At high temperature, in the de-confined world,
the quarks are the correct degrees of freedom in which to measure baryon number and the baryon number is finite.  As one increases the baryon number density at temperatures below the confinement 
temperature, there is a phase which is confined, but the baryon number chemical potential is large
enough that the baryon number density is finite.

The bottom line of these arguments is that there are two order parameters corresponding to 
confinement and to baryon number.  This in principle allows four possible phases.  The phase where
there is de-cofninement and zero baryon number density is apparently not realized in nature, but the other three may be.

We can compute the dependence of the pressure and energy density on $N_c$ for the quarkyonic phase.
To do this, assume the quark chemical potential is large compared to $\Lambda_{QCD}$.  The pressure and energy density are computable in this limit and are of order $N_c$.  It is remarkable that the bulk properties of quarkyonic matter may be computed using perturbation theory at very high density.  This is despite that such matter is confined.  This is not unprecedented, since computations of scattering processes at short distances may be done in perturbation theory, even though the processes take place in the confined vacuum.  The perturbation expansion works for bulk thermodynamic quantities because the major
contribution arises from quarks deep inside the Fermi sea, where short distance interactions dominate.
Pairing processes near the Fermi surface are sensitive to long distance affects, and presumably 
pairs which form near the Fermi surface must be confined.

The fact that the energy density and pressure are proportional to $N_c$ may also be seen in a Skyrmyonic description of baryons for two flavors.  Such a description follows from the large $N_c$ limit.\cite{skyrme}
The action for Skyrmions is
\begin{equation}
   S = \int ~ d^4x ~ \left( f_\pi^2 ~ tr ~V^\mu V^\dagger_\mu + \kappa ~ tr ~[V^\mu,V^\nu]^2
    \right)
\end{equation}
In this equation, $V^\mu$ is a derivative of a n $SU(2)$ group element, and both $f_\pi$ and $\kappa$
are constants of order $N_c$.  We see that the action is therefore of order $N_c$, so that bulk quantities
such as pressure and energy density computed in this theory should also be of order $N_c$

These arguments show that the confined-baryonless phase has bulk properties of order 1 in $N_c$, the de-coniined phase of order $N_c^2$ and the quarkyonic phase is of order $N_c$  The phase diagram for QCD in the large $N_c$ limit is shown in Fig. 3.
\begin{figure}[htbp]
\begin{center}
\includegraphics[width=0.60\textwidth]{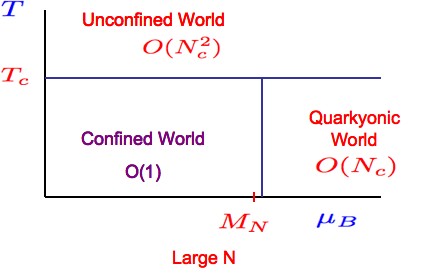} 
\end{center}
\caption{ The phase diagram for large $N_c$ QCD.
}
\end{figure}

The change in the bulk properties of the system when $\mu_B$ crosses the confinement-quarkyonic
transition is of order $N_c$.  We can also estimate the width of the transition region.  Recall that in the large $N_c$ limit baryons are very strongly interacting.  When the number density of baryons becomes
of order $1/\Lambda_{QCD}^3$ or when the Fermi momentum is $k_F \sim \Lambda_{QCD}$,, one will no doubt have made a transition to quarkyonc matter.  This density is controlled by the Fermi momentum.  For baryons with large masses of order $N_c$,
we see that the baryon chemical potential is of order, $\mu_B \sim M_B + k_F^2/2M_B$.  This means that in a width of order $1/N_c$ in $\mu_B$ or of order $1/N_c^2$ in $\mu_Q$,  the transition is achieved.  In the large $N_c$ limit, this width shrinks to zero.

\section{Finite $N_F/N_c$}

The finite ratio of $N_F/N_c$ in the large $N_c$ limit implies that there is no distinction between the confined and de-confined worlds at finite temperature and density.  The baryon number density remains
a valid order parameter.  The transition is driven because of the huge degeneracy of the lowest mass baryon states.  There are of order $e^{N_cG(N_F/N_c)}$ such states, where $G$ is a function determined by Young-Tableau which counts such states.  The baryon  density is
\begin{equation}
   \rho_B \sim e^{N_c G(N_F/N_c)}   e^{-M_B/T + \mu_B/T}
\end{equation}
This has a transition when $\mu_B = M_B -TN_c G(N_c/N_F)$  In this case the world is divided into that
of mesons without baryons, and a world with finite baryon number density.

\section{Phenomenology and Speculation}

It is tempting to speculate on the nature of the phase transition in QCD for $N_c = 3$ and realistic numbers of flavors.  It is difficult a priori to know whether phase transitions remain or whether they become cross overs.  I have little to say about this.  It is clear that the diagram drawn in Fig. 3 becomes smoothed due to finite $N_c$ and $N_F$ effects.  This means that the transition line at small $\mu_B$ and finite $T$ has some small decrease as $\mu_B$ increases.  When $\mu_B \sim  \sqrt{N_c} \Lambda_{QCD}$, the line of finite temperature transitions will have either disappeared or have dipped near $T = 0$.  If it dips down, it probably is a weak transition because the effect of
de-confinement is to liberate gluons and at low $T$ and high $\mu$, and  there the ratio of gluons to quarks is very small.  Most likely, the finite temperature line of transitions ends in a second order point, if it
was ever a first order transition.  Presumably, there is still a tricritical region as in the large $N_c$ limit.
If the transition becomes a cross over, then perhaps a critical point slides along the
line of quarkyonic-confined transition.  

The large $N_c$ considerations must be further developed in order to say anything about the chiral transition.  Chiral symmetry breaking should be a Fermi surface effect in the quarkyonic phase.\cite{austria}  Whether or not it is restored in the quarkyonic phase, or only approximately restored is not resolved.  

A hypothetical phase diagram for QCD is shown in Fig. 4.
\begin{figure}[htbp]
\begin{center}
\includegraphics[width=0.70\textwidth]{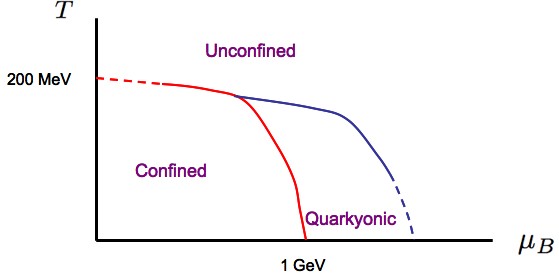} 
\end{center}
\caption{ A hypothetical phase diagram for QCD.
}
\end{figure}

Perhaps the strongest hints for the existence of a quarkyonic phase come from the ratios
of integrated yields of particles produced in heavy ion experiments.\cite{stachel}.  This is
supposed to give information about the decoupling temperature and chemical potential
of matter produced in heavy ion collisions.  Since the energy density jumps by of order
$N_cN_F \Lambda_{QCD}^4$ across the quarkyonic transition, and since in large $N_c$, particle cross sections do not change their $N_c$ dependence\cite{kapusta}, one might expect that the freeze out occurs at the quarkyonic phase transition.

In Figure 5, the supposed decoupling temperature and baryon chemical potential are computed for various energies of experiments.  Note that the line goes to zero temperature at about the nucleon mass.  This is easy to interpret in terms of the quarkyonic phase transition, but difficult to understand
if the line corresponded to the confinement transition.
\begin{figure}[htbp]
\begin{center}
\includegraphics[width=0.60\textwidth]{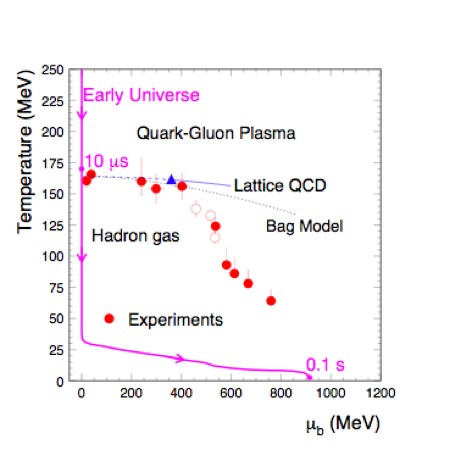} 
\end{center}
\caption{ The decoupling temperature and density of matter produced in heavy ion collisions.
}
\end{figure}

Also shown on the plot are results of computations from the bag model and from lattice gauge theory
which show the weak dependence on baryon chemical potential of the confinement transition line.

There is some lattice data which argues in favor of the existence of the quarkyonic phase.  In the computations of Fodor et. al., micro-canonical techniques were used on lattices of very small size.\cite{fodor} They found the phase diagram shown in Figure 6.
\begin{figure}[htbp]
\begin{center}
\includegraphics[width=0.60\textwidth]{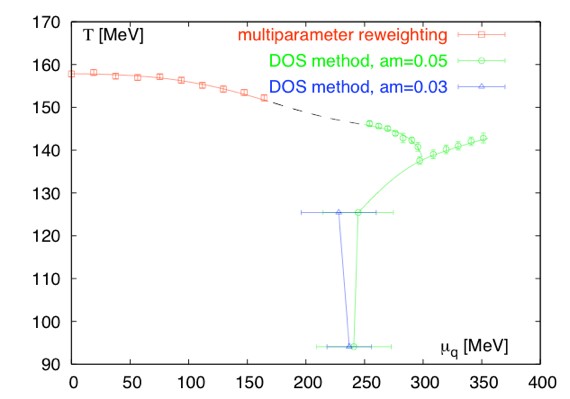} 
\end{center}
\caption{ The phase diagram found by Fodor and colleagues.
}
\end{figure}

\section{Conclusions}

Quarkyonic matter forces us to revise our conception of the phase diagram of QCD.  There are of course many unanswered questions:  How does the chiral transition interplay with the quarkyonic transition?
How is quarkyonic matter related to Skyrmionic crystals?\cite{skyrme}  What is the nature of the Fermi surface of quarkyonic matter?  How is the liquid gas phase transition related to the quarkyonic phase transitions?  These are of course many more.

\section{Acknowledgements}

I thank Arkady Vainshtein and Misha Voloshin for inviting me to Continuous Advances in QCD, where this talk was presented.  Their gracious hospitality is greatly appreciated.  I thank Rob Pisarski and Yoshimasa Hidaka with whom many of these idea were developed.
I thank my colleagues at Rob Pisarski and Yoshimasa Hidaka for their clever insights,
and with whom these ideas were developed.  I also thank Kenji Fukushima for his
intuitive insights concerning such matter, and Kzysztof Redlich for his many insights concerning
the properties of matter at finite baryon number density.
My research is supported under DOE Contract No.
DE-AC02-98CH10886.


\end{document}